**Machine Learning-Guided Screening of Advantageous Solvents for Solid Polymer Electrolytes in Lithium Metal Batteries**

*Jiadong Shen[1], Junjie Chen[1], Xiaosa Xu[1], Jin Li[1], Zhenyu Wang[1], Pengzhu Lin[1], Zixiao Guo[1], Yu Wang[1], Jing Sun[1*], Baoling Huang[1*], Tianshou Zhao[1,2*]*

J. Shen, J. Chen, X. Xu, J. Li, Z. Wang, P. Lin, Z. Guo, Y. Wang, J. Sun, B.L. Huang, T.S. Zhao

[1] Department of Mechanical and Aerospace Engineering, The Hong Kong University of Science and Technology, Clear Water Bay, Kowloon, Hong Kong SAR, China

E-mail: jsunav@connect.ust.hk (J. Sun); mebhuang@ust.hk (B.L. Huang).

[2] Department of Mechanical and Energy Engineering, Southern University of Science and Technology, Shenzhen, 518055, China

E-mail: zhaots@sustech.edu.cn (T.S. Zhao)

## ABSTRACT

Trace residual solvents in solid polymer electrolytes (SPEs) significantly affect electrolyte and interface properties, where optimal selection enhances ionic conductivity and transference numbers. However, solvent complexity hinders general screening methods. We establish a universal criterion linking electronic (HOMO, LUMO) and macroscopic properties (dielectric constant, dipole moment, polarizability) via machine learning on a ~10,000-solvent dataset from high-throughput DFT. Two solvents, *N*-methoxy-*N*-methyl-2,2,2-trifluoroacetamide and 2,2,2-trifluoro-*N*,*N*-dimethylacetamide were identified. Experimental incorporation of trace *N*-methoxy-*N*-methyl-2,2,2-trifluoroacetamide into a poly(vinylidene fluoride-co-hexafluoropropylene) matrix achieves a 4.5 V window, $5.5\times10^{-4}$ S $cm^{-1}$ conductivity (30°C), and 0.78 $Li^+$ transference number. The cell retains 86.7% capacity over 500 cycles ($LiFePO_4$) and 98.7% after 200 cycles at 2C ($LiNi_{0.9}Co_{0.05}Mn_{0.05}O_2$), outperforming 2,2,2-trifluoro-*N,N*-dimethylacetamide, dimethylformamide, *N*-methyl-2-pyrrolidone, and dimethyl sulfoxide. This synergy enables balanced ion transport, wide stability, and cycling durability, advancing safer, high-energy lithium metal batteries. Our integrated approach establishes a solvent screening paradigm for rational SPE design, accelerating next-generation battery development.

The rapid expansion of electric vehicle markets has heightened demands for batteries with greater energy density and improved safety, thereby intensifying interest in solid-state lithium metal batteries.[1-3] Inorganic solid electrolytes, such as oxides, sulfides, and halides, exhibit high ionic conductivity but suffer from significant interfacial resistance due to inadequate solid-solid contact, limiting their commercial application.[4-6] By comparison, solid polymer electrolytes (SPEs) including poly(ethylene oxide), polydioxolane, poly(vinylidene fluoride), and particularly poly(vinylidene fluoride-co-hexafluoropropylene) (PVDF-HFP) have attracted considerable attention for their stable electrode interfaces and favorable processability.[7,8] However, SPEs often have lower ionic conductivity, chemical stability, and mechanical strength than inorganic materials.[9,10] Strategies such as incorporating inorganic fillers can boost mechanical rigidity,[11,12] but introducing organic solvents remains simpler and is widely adopted[13,14] to enhance lithium salt dissociation and stabilize interfaces.[15-21] A typical example is the partial polymerization of dioxolane into polydioxolane,[22] leaving residual monomers that enrich local lithium salt concentrations and improve ion transport. Likewise, solvents like *N,N*-dimethylformamide (DMF), employed to dissolve PVDF-HFP, strongly interact with polymer chains and are difficult to remove completely.[13,15] Rather than discarding such residues as impurities, they can enhance battery performance by creating locally high salt concentrations.[23,24] Hence, clarifying how deliberately added or inadvertently retained solvents affect the performance of SPEs is crucial. Yet, choosing suitable solvents, whether for casting or as performance enhancers, remains challenging.

From a materials science perspective, the highest occupied (HOMO) and the lowest unoccupied (LUMO) molecular orbital levels of solvent molecules strongly influence their macroscopic properties, such as polarizability ($\alpha$), dipole moment ($\mu$), dielectric constant ($\varepsilon$), and donor number (DN)[25,26] which dictate $Li^+$ coordination and transport in solid polymer electrolytes[27]. Leveraging high-throughput DFT and machine learning, we aim to directly link these electronic descriptors to improved ion transport and interfacial stability. This integrated strategy identifies promising solvents for safer, high-performance lithium metal batteries. High polarity can facilitate salt dissociation but may aggravate lithium corrosion.[28] Optimal $\varepsilon$ values (25–30) help balance solvation and interfacial reactions,[26,28] while high DN can improve $Li^+$ coordination at the expense of transport.[29,30] HOMO and LUMO dictate electrochemical stability.[27] Unlike crystalline materials where band structures have guided catalysts and battery electrode designs[31-34], solvent structure–property correlations are less established, often relying on empirical or trial-and-error solvent blending.[35-37] This raises complexity, cost, and compatibility concerns.[38] Machine learning algorithms have recently become powerful tools for material development,[39-43] but face similar drawbacks as earlier high-throughput DFT calculations:[44,45] they remain predominantly theoretical, lack of experimental validation, and rely heavily on data quality, emphasizing the need to integrate experiments with theoretical models.[46-48]

Here, we screened ~10,000 organic solvents containing C, H, O, N, and F to identify beneficial residues for SPEs. High-precision density functional theory (DFT) calculations revealed correlations between HOMO-LUMO gaps and $\mu$, $\varepsilon$, $\alpha$, and DN, verifying the pivotal role of electronic structure in shaping solvent properties. We then employed Extreme Gradient Boosting (XGBoost),[49] Sure Independence Screening and Sparsifying Operator (SISSO),[50] and crystal graph convolutional neural network (CGCNN)[51] to establish a predictive model, proposing criteria for low-polarity solvents. Among these, 2,2,2-trifluoro-*N,N*-dimethylacetamide (TFDMA)[52,53] and *N*-methoxy-*N*-methyl-2,2,2-trifluoroacetamide (TFOMA) emerged as top candidates. TFOMA's additional methoxy group appears to enhance ionic pathways.[26,28,30] Experimentally, trace TFOMA in PVDF-HFP delivers a wider voltage window (>4.5 V), higher ionic conductivity ($5.5 \times 10^{-4}$ S $cm^{-1}$ at 30 °C), and a larger lithium-ion transference number (0.78), along with lower lithium plating/stripping overpotential. In $LiFePO_4$ and $LiNi_{0.9}Co_{0.05}Mn_{0.05}O_2$ (NCM91) systems, PVDF-HFP@TFOMA exhibits excellent cycling stability, retaining 98.9% capacity at 2 C after 200 cycles. This study not only advances SPE research but also demonstrates a data-driven, experimentally validated paradigm for solvent design in lithium metal batteries.

**Screening and Prediction Workflow.** Molecular orbital energy levels (HOMO and LUMO) are key parameters for characterizing the electronic structures of solvent molecules, while properties such as $\alpha$, $\varepsilon$, $\mu$, and *DN* reflect the macroscopic physicochemical properties of solvents.[26] Based on the fundamental principle that electronic structure determines macroscopic properties, we designed a systematic screening strategy for solvent molecules. As shown in **Figure 1**, we extracted organic solvent molecules containing C, H, O, N, and F elements from the ChemSpider database[54] and obtained basic physical parameters, including HOMO and LUMO energy levels, $\alpha$, and $\mu$, through high-throughput DFT calculations. The $\varepsilon$ were calculated using the Clausius-Mossotti equation[55], and the *DN* were determined by calculating the interaction energies between solvent molecules and the Lewis acid $SbCl_5$.[29,56] We first established structure–property relationships between molecular orbital energy levels (HOMO-LUMO gap) and $\varepsilon$ as well as $\mu$. On this basis, further DFT calculations were conducted to explore the interfacial properties of solvents, including *DN* and Li binding energies. Subsequently, based on the dataset obtained from high-throughput calculations, we employed machine learning algorithms to deeply reveal the intrinsic connections between electronic structures and macroscopic properties, and experimentally validated the relevant predictions. The detailed screening and prediction process is illustrated in **Figure S1**.

**High-Throughput DFT Calculations**. Building upon our systematic screening strategy, we analyzed the relationships among the HOMO-LUMO gaps, $\mu$, and $\varepsilon$ of the solvent molecules obtained from high-throughput DFT calculations. These relationships are presented in **Figure 2**. Additionally, the correlations between the relative polarizabilities of the molecules and the solvents' $\varepsilon$, $\mu$, and HOMO-LUMO gaps are depicted in **Figure S2**. As shown in **Figure 2a**, the HOMO-LUMO gaps of the solvent molecules are primarily concentrated between 2–6 eV, the $\mu$ range from 0 to 15 Debye (D), and the $\varepsilon$ are predominantly between 22 and 30. **Figure 2b** illustrates a significant positive correlation (slope $\approx$ 0.11) between the $\mu$ and $\varepsilon$ of the solvent molecules, indicating that molecules with larger $\mu$ tend to exhibit higher $\varepsilon$. Similar positive correlations between both isotropic and anisotropic relative $\alpha$ and the $\varepsilon$ (**Figures S2a** and **S2b**) are revealed in further analysis. This fundamental relationship stems from the contributions of deformation polarizability and orientation polarizability to the dielectric constant (**Supplementary note (1)**), where the Clausius–Mosotti equation and Debye's theory jointly rationalize how $\mu$ and $\alpha$ determine $\varepsilon$. This correlation is also evident in the relationships between relative polarizability and $\mu$ (**Figures S2c** and **S2d**). These findings suggest a universal correlation between the microscopic structural characteristics of solvent molecules and their macroscopic dielectric properties: larger $\mu$ and relative polarizabilities generally correspond to higher $\varepsilon$.

Moreover, we investigated how the $\mu$ and $\varepsilon$ of the molecules vary with their electronic structures, as shown in **Figures 2c** and **2d**. It can be observed that within the HOMO-LUMO gap range of 0.5 eV to 2 eV, the $\mu$ and $\varepsilon$ of the solvents decrease sharply as the gap increases. This trend becomes significantly less pronounced when the gap further increases to the range of 2–8.5 eV. Such a nonlinear dependence on the HOMO-LUMO gap can be understood via second-order perturbation theory, wherein polarizability exhibits a strong inverse correlation with the gap (**Supplementary note (2)**), thus explaining the sharp decline in $\mu$ and $\varepsilon$ when the gap is small. As the gap increases, the charge distribution within the molecule becomes more localized, leading to a gradual decrease in polarity[57, 58]. A similar phenomenon is observed in the polarizabilities of the molecules, as shown in **Figures S2e** and **S2f**. These insights, derived from correlating various physical properties of solvent molecules, provide a new perspective for understanding the relationship between the electronic structure and properties of solvents. These correlations emphasize that tuning HOMO/LUMO energies, dipole moment, and dielectric constant is essential for balancing $Li^+$ solvation and chemical stability in polymer electrolytes. Such understanding offers important guidance for the rational design of electrolyte solvents in lithium-ion batteries.

**Construction and Validation of the Machine Learning Model**. To develop our predictive framework, we first assembled a dataset of solvents and their key properties (including dielectric constant ($\varepsilon$), dipole moment ($\mu$), and HOMO-LUMO energy levels) from the ChemSpider database[54], a resource that provides verified chemical structures and physicochemical data. By leveraging ChemSpider, we ensured comprehensive coverage of diverse solvents to support subsequent property calculations and feature extraction. Building on this foundation, we then employed a multi-level machine learning strategy encompassing XGBoost,[49] SISSO,[50] and CGCNN[14] aiming to balance interpretability and predictive accuracy. Specifically, XGBoost and SISSO were selected because they provide relatively straightforward interpretation of feature importance, making them well-suited for quickly screening large numbers of descriptors. Meanwhile, CGCNN excels at capturing complex non-linear interactions, particularly for periodic structures, thus extending predictive capabilities to systems where local atomic environments significantly influence solvent properties. Initially, we used interpretable algorithms such as XGBoost[49] and SISSO[50] as screening tools to identify critical features and assess predictive performance. As illustrated in **Figure S3a**, neglecting HOMO-LUMO descriptors resulted in poor accuracy for predicting *ε* ($R^2 = 0.19$). Incorporating these orbital-level descriptors significantly enhanced the model ($R^2 = 0.60$, **Figure S3b**), with feature importance analysis (**Figure S3c**) revealing that electronic structure attributes contributed over 65%. SISSO corroborated these findings: using the HOMO-LUMO gap alone yielded $R^2 = 0.65$ (**Figure S3d**), while additional parameters (molecular weight, van der Waals volume) offered minimal improvement ($\Delta R^2 < 0.02$, **Figures S3e,f**). Conversely, models relying purely on geometric features (*e.g.*, bond lengths, dihedral angles) performed poorly ($R^2 < 0.1$, **Figure S3g**), underscoring the decisive role of electronic structure in shaping solvent properties.

Despite these advances, highly nonlinear properties such as *μ* remained challenging for XGBoost and SISSO (**Figures S3h,i**). To address this limitation, we implemented the **CGCNN**[51] algorithm, originally designed for crystalline materials, by representing each solvent molecule under periodic boundary conditions (**Figure 3a and Figure S4**). In this approach, atoms and bonds become nodes and edges in a graph network, enabling the capture of both intra- and intermolecular interactions. Through multiple convolutional, hidden, and pooling layers, the CGCNN automatically learns hierarchical structural and electronic features. Remarkably, it achieved $R^2 = 0.91$ for *μ*, 0.97 for *ε*, and 0.98 for the HOMO-LUMO gap (**Figures 3b–d**). To further explore the "black box" nature of deep learning, we performed dimensionality reduction analyses, revealing strong correlations between the CGCNN's latent embeddings and electronic-structure descriptors (**Table S1**). Notably, the predicted outputs shifted from strong alignment with LUMO levels in one subset ($r = 0.937$, $p < 10^{-30}$) to closer correspondence with HOMO energies in another ($r = -0.568$, $p < 10^{-30}$), illustrating the model's adaptability to distinct quantum-chemical contexts. Further principal component and *t*-SNE evaluations showed stable correlations between latent coordinates and dipole moments (*e.g.*, PC1 vs dipole: $r = -0.506$, $p < 10^{-30}$), indicating that these internal representations capture crucial molecular orbital and electron density features. The SHAP contribution analysis (**Figures S5–S7**) clarifies how each descriptor influences the predicted properties. Specifically, **Figures S7a** and **S7b**, which were derived from the XGBoost models, highlight the dominant role of frontier orbitals (HOMO, LUMO) in determining both the dielectric constant and dipole moment. This finding is consistent with the known importance of electronic structure in polarization response and charge redistribution. The positive SHAP values for higher-energy HOMOs suggest stronger dielectric effects through easier electron donation, while lower LUMO energies enhance molecular polarizability. In addition, total polarizability and electronegative-atom fractions further underscore the impact of molecular asymmetry on dipole moments. In contrast, **Figures S7c** and **S7d**, which were obtained from the CGCNN models, show that latent embeddings rank highly for the HOMO–LUMO gap, capturing deeper structural–electronic correlations beyond simpler atomic descriptors. Their consistent relevance confirms alignment with fundamental quantum-chemical factors and reinforces model interpretability. Overall, these results demonstrate that the chosen descriptors, along with the latent GNN features, accurately mirror known theoretical dependencies on orbital energies and polarizability, thereby lending confidence to the robustness and transparency of our approach. Further confirming its robustness, we also carried out 50-fold cross-validation, multiple-seed tests, and y-scrambling (**Figure S8**), which demonstrated stable prediction accuracy and supported genuine structure–property learning rather than mere data memorization. Overall, this integrated approach provides a solid theoretical basis for applying deep neural networks in molecular design, linking electronic structures to macroscopic solvent properties.

**Establishing and Applying Solvent Screening Criteria for Novel Solvents**. Building on our systematic solvent analysis (**Table S2**), we defined quantitative criteria (**Figure 4a**) to identify ideal candidates: HOMO < –6 eV, LUMO > –1 eV, a HOMO–LUMO gap > 5 eV, $\varepsilon$ in the 20–50 range, and $\mu$ between 3–6 D. Such parameters balance oxidative/reductive stability (via HOMO and LUMO), sufficient salt dissociation ($\varepsilon$), and moderated ion–solvent interactions ($\mu$).[27,29,37] Through statistical ranking of molecules that met these benchmarks, the top ten solvents were identified, with five leading contenders ($C_4F_6H_3NO$, $C_4F_6H_3NO_2$, $C_6F_{10}H_3NO$, $C_6FH_8NO_2$, $C_8FH_{12}NO_2$) shown in **Figure 4b** (the remaining five are shown in **Figure S9**). Their interactions with LiTFSI, $SbCl_5$, and Li were examined (**Figure 4c**, **Figures S10–S12**), using $SbCl_5$ adsorption energies to quantify DN.[29] As illustrated in **Figure S13**, DN positively correlates with the HOMO–LUMO gap, indicating stronger electron donation for wider gaps. However, solvents with higher DN often exhibit weaker $Li^+$ acceptance, reflecting a trade-off in coordinating lithium ions.

Expanding on this insight (**Figure 4d**), we identified two promising, previously untrained solvents: 2,2,2-trifluoro-*N,N*-dimethylacetamide (TFDMA) ($C_4H_6F_3NO$) and *N*-methoxy-*N*-methyl-2,2,2-trifluoroacetamide (TFOMA) ($C_4H_6F_3NO_2$) (**Figure 4e**). TFDMA has been experimentally validated,[48] while TFOMA is proposed in this study for the first time. TFOMA's HOMO (–7.10, –6.77 eV) and LUMO (–0.47, –0.26 eV) suggest enhanced high-voltage stability and favorable SEI formation. By contrast, TFDMA's apparently positive LUMO (2.06, 1.99 eV) may reflect computational parameter differences and thus warrants caution. MD simulations further support TFOMA's superior interfacial behavior and ionic conductivity (**Figure 4f**, **Table S3 and Table S4**).

In **Figure 4g**, TFOMA exhibits localized high-concentration structures with lithium salts, while $Li^+$ root mean square displacement (**Figure 4h**) is higher for TFOMA (0.01 Å $ns^{-1}$) than TFDMA (0.007 Å $ns^{-1}$). Additional simulations at 300–340 K (**Figures S14a–d**) confirm accelerated $Li^+$ transport. The radial distribution function (**Figure 4i**) reveals longer C–F distances and more concentrated distributions in TFOMA, implying a distinct solvation mechanism (also see **Figures S14e,f**). We attribute this to the extra methoxy group in TFOMA, which forms new ion channels, thereby boosting macroscopic conductivity. This emphasizes the importance of functional group engineering for optimizing electrolyte performance.

**Electrochemical Performance Testing and Mechanistic Analysis of Solid-State Lithium Metal Full Cells**. To validate our design paradigm, we fabricated PVDF-HFP-based solid polymer electrolytes (SPEs) with TFOMA or TFDMA as the residual solvent. PVDF-HFP was selected for its excellent high-voltage stability and strong F···F interactions with fluorinated solvents,19,48 which promote a uniform solvent distribution and retain beneficial traces of solvent for enhanced lithium-ion conduction. Elemental mapping confirmed a uniform distribution of C, N, O, F, and S in membranes of ~23 μm thickness (**Figures S15, S16**). A trace (~0.1 μL) of each solvent was introduced during cell assembly to ensure consistent residual content. Linear sweep voltammetry (**Figure 5a**) reveals that PVDF-HFP@TFOMA can withstand oxidation up to 4.5 V, whereas PVDF-HFP@TFDMA oxidizes at 4.25 V. The improved oxidative stability in TFOMA-based electrolytes is attributed to TFOMA's lower HOMO energy and reduced polarity, which collectively suppress unwanted cathode–solvent interactions. At 30 °C, ionic conductivity measurements (**Figure 5b**) show a marked difference: $5.5 \times 10^{-4}$ S $cm^{-1}$ for TFOMA versus $1.0 \times 10^{-4}$ S $cm^{-1}$ for TFDMA. Above 70 °C, the TFOMA system exceeds $10^{-3}$ S $cm^{-1}$, underscoring its superior ion transport properties. Arrhenius fits yield activation energies of 0.29 eV (TFOMA) and 0.45 eV (TFDMA) (**Figure S17**), indicating more efficient $Li^+$ migration with TFOMA. The methoxy moiety in TFOMA likely creates additional ion coordination sites, reducing ion–anion coulombic interactions and fostering higher lithium-ion transference (0.78 *vs* 0.75, **Figure S18**).

In lithium metal symmetric cells at 0.1–0.5 mA $cm^{-2}$, the TFOMA-based electrolyte achieves plating/stripping overpotentials of ±0.025–0.08 V, significantly lower than ±0.05–0.19 V for TFDMA (**Figure 5c**). When paired with a $LiFePO_4$ cathode (2 C at 30 °C), PVDF-HFP@TFOMA delivers 144 mAh $g^{-1}$ versus PVDF-HFP@TFDMA's 130 mAh $g^{-1}$ (**Figure 5d**). Subsequent rate tests yield capacities of 172, 170, 162, 144, and 170 mAh $g^{-1}$ upon cycling at 0.1, 0.2, 1, 2, and back to 0.2 C, respectively (**Figure 5e**). Over 500 cycles at 2 C, capacity retention remains 86.7%, exceeding TFDMA (83.3%) with a Coulombic efficiency near 99.1% (**Figures 5f, S19**). Furthermore, under a high-voltage $LiNi_{0.9}Co_{0.05}Mn_{0.05}O_2$ (NCM91) cathode, TFOMA demonstrates even better performance, achieving 198 and 175 mAh $g^{-1}$ at 0.2 and 0.5 C (**Figure 5g**) and retaining 98.7% of its capacity after 200 cycles at 2 C (**Figure 5i**). Such robust cycling confirms the practical advantage of TFOMA in high-voltage settings, as summarized in **Table S5**, highlighting our data-driven solvent design strategy for next-generation lithium metal batteries. Meanwhile, **Table S6** shows that previous ML-based solvent-screening studies typically focused on narrower property sets with limited validation, whereas our integrated approach broadens predictive coverage and offers deeper experimental verification for advanced SPE development.

In conclusion, we developed a comprehensive strategy for designing and screening solvents for SPEs in lithium metal batteries. High-throughput DFT calculations covering nearly 10,000 candidates uncovered key correlations between HOMO and LUMO levels and macroscopic properties (α, μ, ε,

DN). Using machine learning (XGBoost, SISSO, CGCNN), we built predictive models, established universal screening criteria, and identified five promising solvents. Among them, the newly proposed TFOMA, with its lower polarity and reduced Li-ion interactions, outperformed PVDF-HFP@TFDMA with a 4.5 V electrochemical window, an ionic conductivity of $5.5 \times 10^{-4}$ S $cm^{-1}$ (30 °C), and enhanced Li-metal stability. Cells with PVDF-HFP@TFOMA retained 86.7% of their capacity after 500 cycles using $LiFePO_4$ and 98.7% after 200 cycles at 2 C with NCM91. Our integrated approach merges first-principles theory, machine learning, and experimental validation, accelerating materials discovery and deepening our understanding of solvents' electronic structure–property relationships. This data-driven method can be extended to other advanced electrolyte designs for high-performance, safe solid-state lithium metal batteries.

**ASSOCIATED CONTENT**

Supporting Information: Detailed high-throughput calculation protocols, machine learning model construction and validation procedures, experimental methods, electrochemical testing and characterization data, and additional figures and tables,

**CODE AND DATA AVAILABILITY**

All code and models are publicly available at: https://github.com/mejiadongs. Training datasets can be requested from the authors with valid research justification.

**AUTHOR INFORMATION**

**Author Contributions**

T. Zhao, B. Huang, J. Sun and J. Shen conceived the idea and designed the project. J. Shen designed theoretical models and computational study. J. Shen implemented and constructed machine learning models. J. Shen conducted DFT and MD calculations. J. Shen performed the materials synthesis and carried out the electrochemical experiments. B. Huang, J. Sun, and T. Zhao supervised the project. J. Shen wrote the first draft of the paper. All authors analyzed the data and discussed the results and commented on the manuscript.

## NOTES

The authors declare no conflict of interest.

## ACKNOWLEDGMENTS

This work was fully supported by a grant from the Research Grants Council of the Hong Kong Special Administrative Region, China (Project No. R6005-20). We would also like to extend our thanks to the High-Performance Computing Cluster (HPC3) of the Hong Kong University of Science and Technology for providing the computational resources necessary for this research.

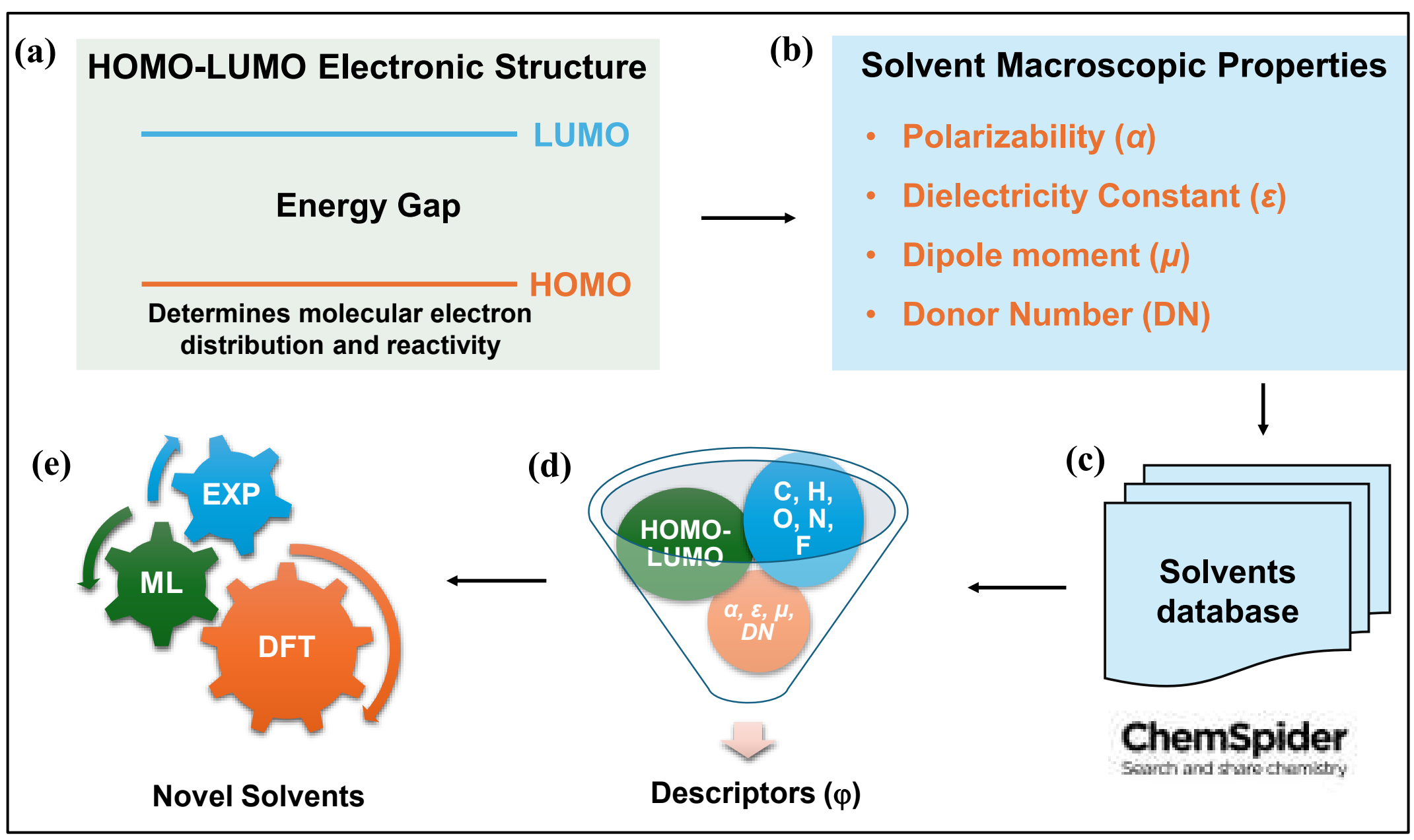

**Figure 1.** (a, b) Starting from the potential relationship between the electronic structure of organic compounds' HOMO-LUMO (a) and corresponding solvent properties ($\alpha$, $\varepsilon$, $\mu$, and *DN*) (b). Data extraction was performed by connecting with the Chemical Spider database (c). (d) Elements were restricted to C, H, O, N, and F to build relationships between HOMO-LUMO, and macroscopic solvent properties, leading to the development of descriptors. (e) Based on DFT results, a machine learning model was trained to predict solvents, followed by experimental validation of the predicted solvents.

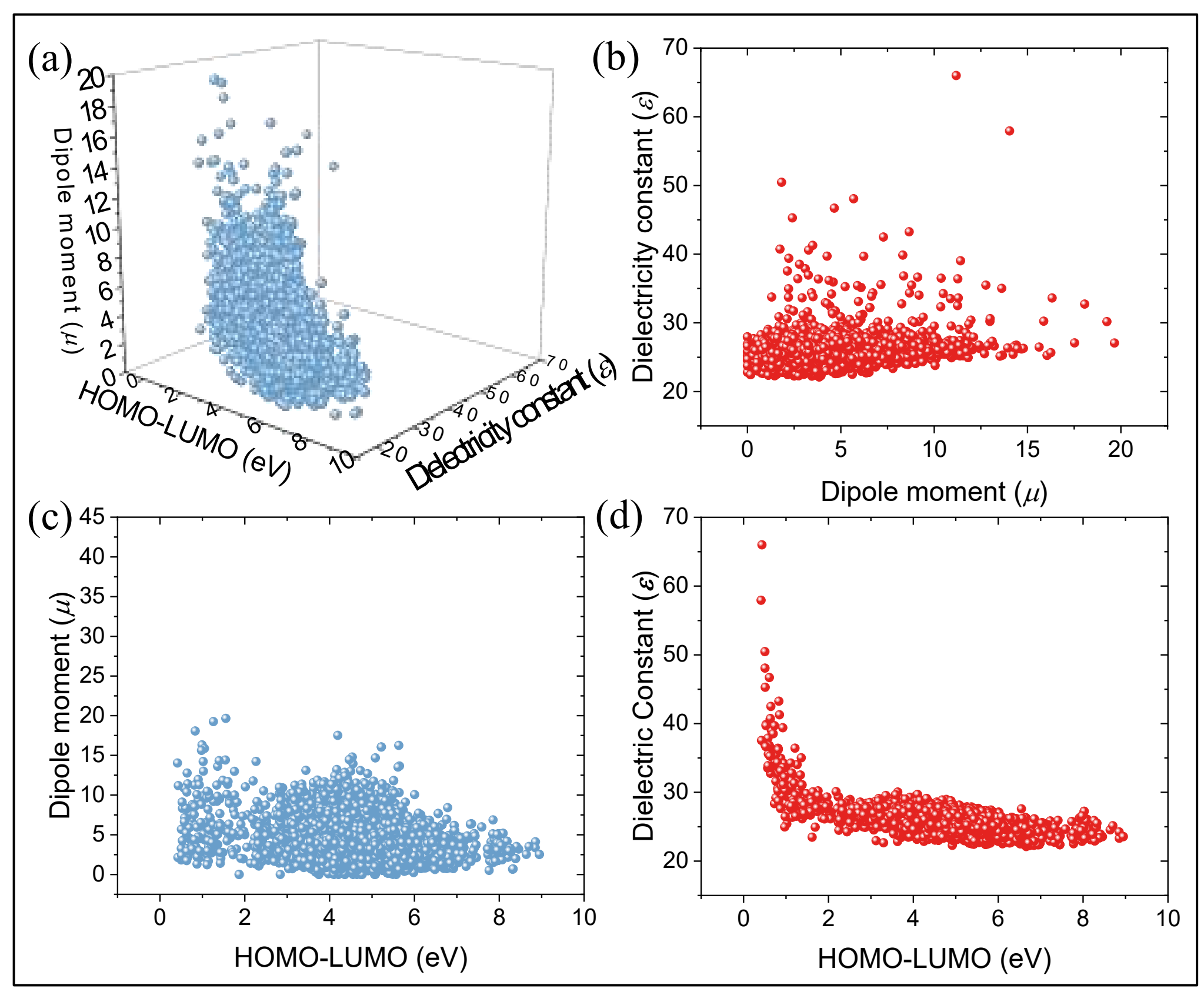

**Figure 2.** (a) 3D scatter plot showing relationships among HOMO-LUMO gap, dipole moment ($\mu$), and dielectric constant ($\varepsilon$); (b) dipole moment ($\mu$) *vs.* dielectric constant ($\varepsilon$); (c) dipole moment ($\mu$) *vs.* HOMO-LUMO gap; (d) dielectric constant ($\varepsilon$) *vs.* HOMO-LUMO gap.

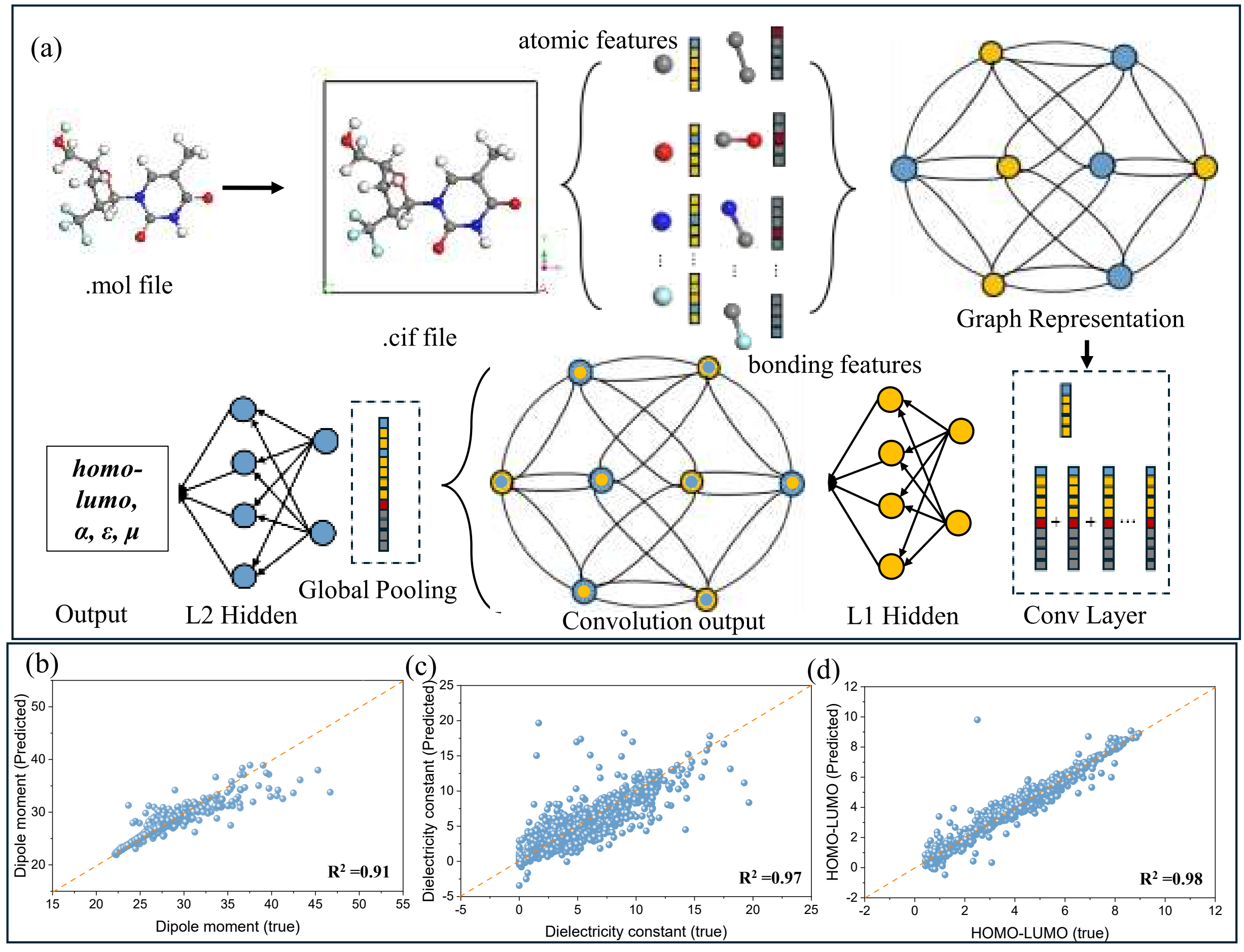

**Figure 3.** (a) Workflow of the CGCNN model construction. (b-d) Prediction results on the test dataset for (b) dielectric constant ($\varepsilon$), (c) dipole moment ($\mu$), and (d) HOMO-LUMO.

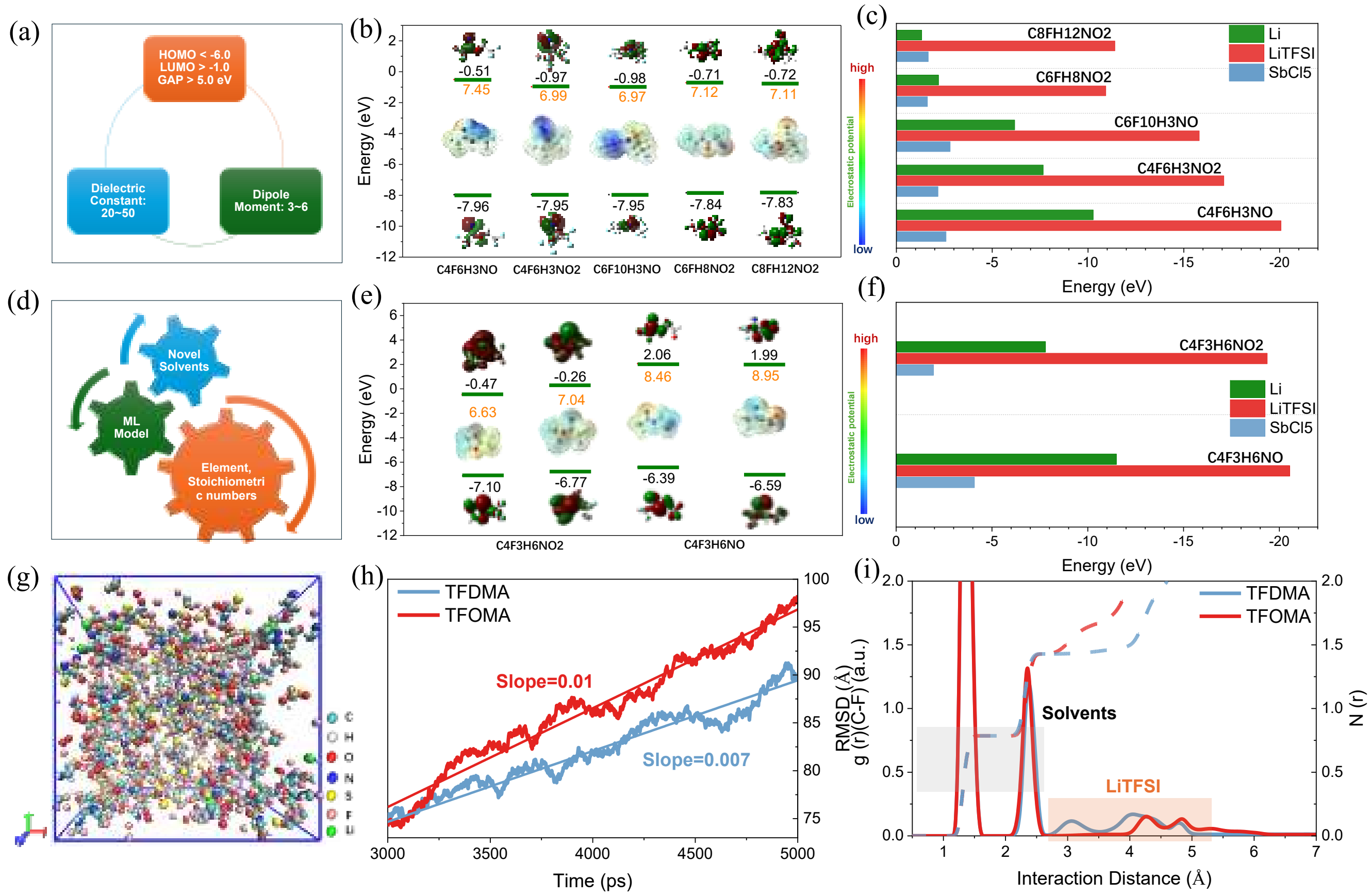

**Figure 4.** (a) Solvent screening criteria; (b) HOMO-LUMO results and molecular structures for the top 5 candidate solvents meeting the criteria; (c) Binding energies of the top 5 solvents with typical lithium salt (LiTFSI) and a representative Lewis acid ($SbCl_5$), including Li binding energy; (d) Prediction of novel solvent properties using element-based features and trained machine learning models; (e) HOMO-LUMO results and molecular structures of two predicted novel solvents; (f) Binding energies between the predicted solvents and typical LiTFSI and $SbCl_5$. (g) Schematic of the TFOMA solvent with a local high concentrated LiTFSI salt model; (h) 5 ns RMSD curves of lithium ions in TFDMA and TFOMA in an equilibrated system from molecular dynamic simulation; (i) Radial distribution functions (RDFs) of C-F atomic pairs associated with LiTFSI and solvent molecules in TFOMA and TFDMA.

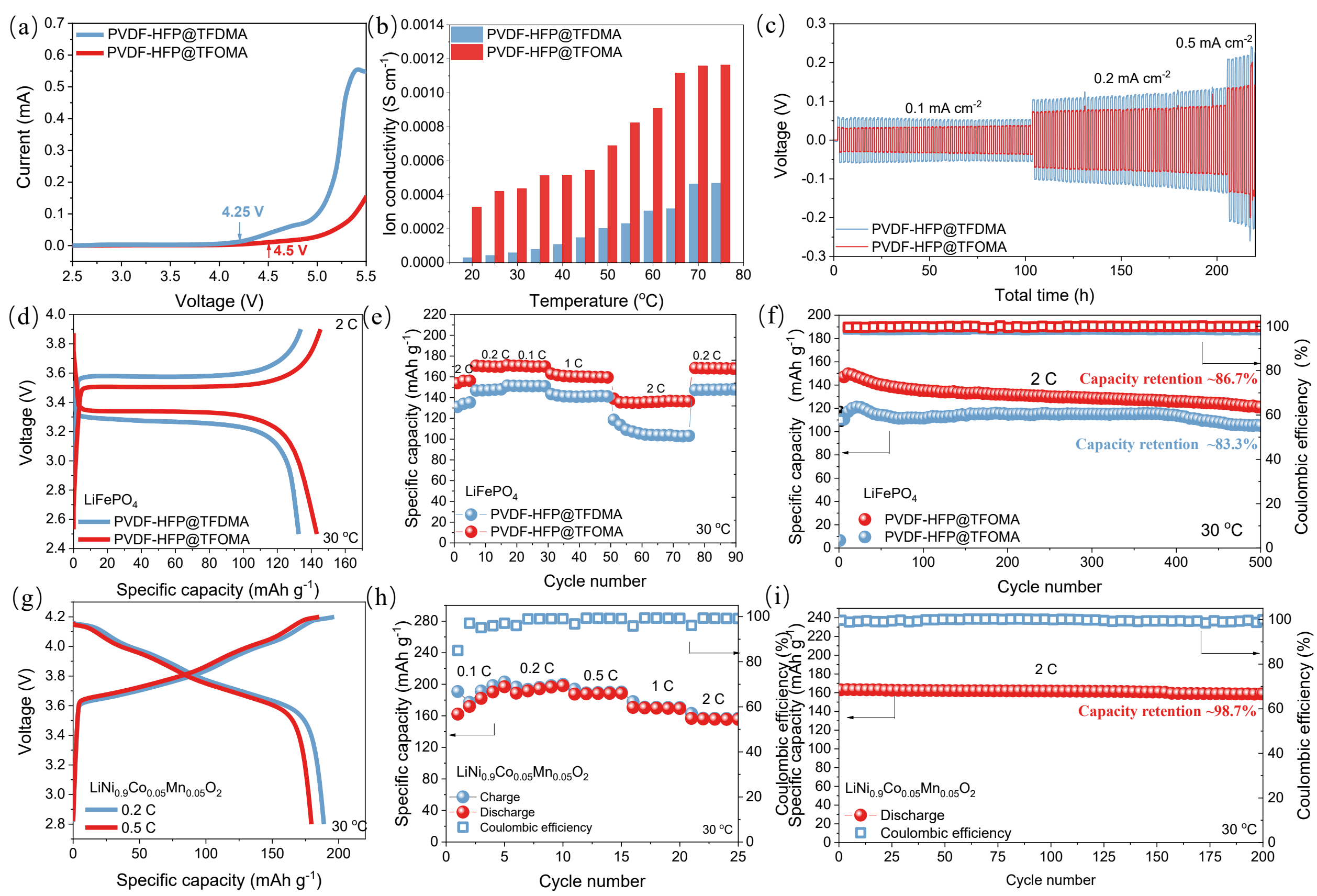

**Figure 5.** (a-c) Comparison of LSV test curves (a), lithium-ion conductivity results (b), and deposition/stripping electrochemical performance (c) between PVDF-HFP@TFDMA and PVDF-HFP@TFOMA solid electrolytes. (d-f) Charge-discharge curves (d), rate performance (e), and long-cycle stability at 2 C (1 C = 175 mAh $g^{-1}$) (f) of $LiFePO_4$ cathode with PVDF-HFP@TFDMA and PVDF-HFP@TFOMA solid electrolytes. (g-i) Charge-discharge curve (g), rate performance (h), and electrochemical performance at 2 C (1 C = 223 mAh $g^{-1}$) (i) of $LiNi_{0.9}Co_{0.05}Mn_{0.05}O_2$ cathode with PVDF-HFP@TFOMA solid electrolyte.

## References

(1) Chen, R.; Li, Q.; Yu, X.; Chen, L.; Li, H. Approaching Practically Accessible Solid-State Batteries: Stability Issues Related to Solid Electrolytes and Interfaces. *Chem. Rev.* **2020**, *120*, 6820–6877.

(2) Ma, J.; Li, Y.; Grundish, N. S.; Goodenough, J. B.; Chen, Y.; Guo, L.; Peng, Z.; Qi, X.; Yang, F.; Qie, L.; Wang, C.; Huang, B.; Huang, Z.; Chen, L.; Su, D.; Wang, G.; Peng, X.; Chen, Z.; Yang, J.; He, S.; Zhang, X.; Yu, H.; Fu, C.; Jiang, M.; Deng, W.; Sun, C.; Pan, Q.; Tang, Y.; Li, X.; Ji, X.; Wan, F.; Niu, Z.; Lian, F.; Wang, C.; Wallace, G. G.; Fan, M.; Meng, Q.; Xin, S.; Guo, Y.; Wan, L. The 2021 battery technology roadmap. *J. Phys. D: Appl. Phys.* **2021**, *54*, 183001.

(3) Wang, C.; Kim, J. T.; Wang, C.; Sun, X. Progress and Prospects of Inorganic Solid-State Electrolyte-Based All-Solid-State Pouch Cells. *Adv. Mater.* **2023**, *35*, 2209074

(4) Lu, Y.; Li, L.; Zhang, Q.; Niu, Z.; Chen, J. Electrolyte and Interface Engineering for Solid-State Sodium Batteries. *Joule* **2018**, *2*, 1747–1770.

(5) Gao, Z.; Sun, H.; Fu, L.; Ye, F.; Zhang, Y.; Luo, W.; Huang, Y. Promises, Challenges, and Recent Progress of Inorganic Solid-State Electrolytes for All-Solid-State Lithium Batteries. *Adv. Mater.* **2018***, 30*, 1705702.

(6) Liu, H.; Cheng, X.; Huang, J.; Yuan, H.; Lu, Y.; Yan, C.; Zhu, G.; Xu, R.; Zhao, C.; Hou, L.; He, C.; Kaskel, S.; Zhang, Q. Controlling Dendrite Growth in Solid-State Electrolytes. *ACS Energy Lett.* **2020***, 5*, 833–843.

(7) Zhao, Q.; Stalin, S.; Zhao, C.; Archer, L. A. Designing solid-state electrolytes for safe, energy-dense batteries. *Nat Rev Mater* **2020***, 5*, 229–252.

(8) Liu, S.; Liu, W.; Ba, D.; Zhao, Y.; Ye, Y.; Li, Y.; Liu, J. Filler-Integrated Composite Polymer Electrolyte for Solid-State Lithium Batteries. *Adv. Mater.* **2023***, 35*, 2110423

(9) Fan, L.; He, H.; Nan, C. Tailoring inorganic–polymer composites for the mass production of solid-state batteries. *Nat Rev Mater* **2021***, 6*, 1003–1019.

(10) Ding, P.; Lin, Z.; Guo, X.; Wu, L.; Wang, Y.; Guo, H.; Li, L.; Yu, H. Polymer electrolytes and interfaces in solid-state lithium metal batteries. *Materials Today* **2021***, 51*, 449–474.

(11) Gong, W.; Ouyang, Y.; Guo, S.; Xiao, Y.; Zeng, Q.; Li, D.; Xie, Y.; Zhang, Q.; Huang, S. Covalent Organic Framework with Multi-Cationic Molecular Chains for Gate Mechanism Controlled Superionic Conduction in All-Solid-State Batteries. *Angew. Chem. -Int. Edit.* **2023***, 62*, 202302505.

(12) Li, B.; Wang, C.; Yu, R.; Han, J.; Jiang, S.; Zhang, C.; He, S. Recent progress on metal–organic framework/polymer composite electrolytes for solid-state lithium metal batteries: ion transport regulation and interface engineering. *Energy Environ. Sci.* **2024***, 17*, 1854–1884.

(13) Zou, S.; Yang, Y.; Wang, J.; Zhou, X.; Wan, X.; Zhu, M.; Liu, J. In situ polymerization of solid-state polymer electrolytes for lithium metal batteries: a review. *Energy & Environmental Science* **2024***, 17*, 4426–4460.

(14) Liu, Y.; Zhang, D.; Luo, L.; Li, Z.; Lin, H.; Liu, J.; Zhao, Y.; Hu, R.; Zhu, M. Percolating coordinated ion transport cells in polymer electrolytes to realize room-temperature solid-state lithium metal batteries. *Energy Storage Materials* **2024***, 70*, 103548.

(15) Vijayakumar, V.; Anothumakkool, B.; Kurungot, S.; Winter, M.; Nair, J. R. *In situ* polymerization process: an essential design tool for lithium polymer batteries. *Energy Environ. Sci.* **2021***, 14*, 2708–2788.

(16) He, K.; Cheng, S. H.; Hu, J.; Zhang, Y.; Yang, H.; Liu, Y.; Liao, W.; Chen, D.; Liao, C.; Cheng, X.; Lu, Z.; He, J.; Tang, J.; Li, R. K. Y.; Liu, C. In-Situ Intermolecular Interaction in Composite Polymer Electrolyte for Ultralong Life Quasi-Solid-State Lithium Metal Batteries. *Angew. Chem. -Int. Edit.* **2021***, 60*, 12116–12123.

(17) Wang, Z.; Zhao, C.; Sun, S.; Liu, Y.; Wang, Z.; Li, S.; Zhang, R.; Yuan, H.; Huang, J. Achieving high-energy and high-safety lithium metal batteries with high-voltage-stable solid electrolytes. *Matter* **2023***, 6*, 1096–1124.

(18) An, H.; Li, M.; Liu, Q.; Song, Y.; Liu, J.; Yu, Z.; Liu, X.; Deng, B.; Wang, J. Strong Lewis-acid coordinated PEO electrolyte achieves 4.8 V-class all-solid-state batteries over 580 Wh kg−1. *Nat Commun* **2024***, 15*, 9150.

(19) Yang, W.; Liu, Y.; Sun, X.; He, Z.; He, P.; Zhou, H. Solvation-Tailored PVDF-Based Solid-State Electrolyte for High-Voltage Lithium Metal Batteries. *Angew. Chem. -Int. Edit.* **2024***, 63*, 202401428.

(20) Zhao, Z.; Zhou, X.; Zhang, B.; Huang, F.; Wang, Y.; Ma, Z.; Liu, J. Regulating Steric Hindrance of Porous Organic Polymers in Composite Solid-State Electrolytes to Induce the Formation of LiF-Rich SEI in Li-Ion Batteries. *Angew. Chem. -Int. Edit.* **2023***, 62*, 202308738.

(21) Liu, Y.; Hu, R.; Zhang, D.; Liu, J.; Liu, F.; Cui, J.; Lin, Z.; Wu, J.; Zhu, M. Constructing Li-Rich Artificial SEI Layer in Alloy-Polymer Composite Electrolyte to Achieve High Ionic Conductivity for All Solid-State Lithium Metal Batteries. *Adv. Mater.* **2021***, 33*, 2004711.

(22) Zhao, Q.; Liu, X.; Stalin, S.; Khan, K.; Archer, L. A. Solid-state polymer electrolytes with in-built fast interfacial transport for secondary lithium batteries. *Nat Energy* **2019***, 4*, 365–373.

(23) Zhang, J.; Xu, W.; Xiao, J.; Cao, X.; Liu, J. Lithium Metal Anodes with Nonaqueous Electrolytes. *Chem. Rev.* **2020***, 120*, 13312–13348.

(24) Efaw, C. M.; Wu, Q.; Gao, N.; Zhang, Y.; Zhu, H.; Gering, K.; Hurley, M. F.; Xiong, H.; Hu, E.; Cao, X.; Xu, W.; Zhang, J.; Dufek, E. J.; Xiao, J.; Yang, X.; Liu, J.; Qi, Y.; Li, B. Localized high-concentration electrolytes get more localized through micelle-like structures. *Nat. Mater.* **2023**, *22*, 1531–1539.

(25) Liu, Q.; Wang, L. Fundamentals of Electrolyte Design for Wide-Temperature Lithium Metal Batteries. *Adv. Energy Mater.* **2023***, 13*, 2301742.

(26) Li, M.; Hicks, R. P.; Chen, Z.; Luo, C.; Guo, J.; Wang, C.; Xu, Y. Electrolytes in Organic Batteries. *Chem. Rev.* **2023***, 123*, 1712–1773.

(27) Peljo, P.; Girault, H. H. Electrochemical potential window of battery electrolytes: the HOMO–LUMO misconception. *Energy Environ. Sci.* **2018***, 11*, 2306–2309.

(28) Wang, Z.; Sun, Z.; Li, J.; Shi, Y.; Sun, C.; An, B.; Cheng, H.; Li, F. Insights into the deposition chemistry of Li ions in nonaqueous electrolyte for stable Li anodes. *Chem. Soc. Rev.* **2021***, 50*, 3178–3210.

(29) Zhou, P.; Xiang, Y.; Liu, K. Understanding and applying the donor number of electrolytes in lithium metal batteries. *Energy Environ. Sci.* **2024***, 17*, 8057–8077.

(30) Wan, S.; Ma, W.; Wang, Y.; Xiao, Y.; Chen, S. Electrolytes Design for Extending the Temperature Adaptability of Lithium-Ion Batteries: from Fundamentals to Strategies. *Adv. Mater.* **2024***, 36*. 2311912.

(31) Seh, Z. W.; Kibsgaard, J.; Dickens, C. F.; Chorkendorff, I.; Nørskov, J. K.; Jaramillo, T. F. Combining theory and experiment in electrocatalysis: Insights into materials design. *Science* **2017***, 355*, eaad4998.

(32) Shen, J.; Xu, X.; Liu, J.; Wang, Z.; Zuo, S.; Liu, Z.; Zhang, D.; Liu, J.; Zhu, M. Unraveling the Catalytic Activity of Fe-Based Compounds toward $Li_2S_x$ in Li-S Chemical System from *d-p* Bands. *Adv. Energy Mater.* **2021***, 11*, 2100673.

(33) Zhou, J.; Liu, X.; Zhu, L.; Zhou, J.; Guan, Y.; Chen, L.; Niu, S.; Cai, J.; Sun, D.; Zhu, Y.; Du, J.; Wang, G.; Qian, Y. Deciphering the Modulation Essence of p Bands in Co-Based Compounds on Li-S Chemistry. *Joule* **2018***, 2*, 2681–2693.

(34) Shen, J.; Liang, Z.; Gu, T.; Sun, Z.; Wu, Y.; Liu, X.; Liu, J.; Zhang, X.; Liu, J.; Shen, L.; Zhu, M.; Liu, J. Revisiting the unified principle for single-atom electrocatalysts in the sulfur reduction reaction: from liquid to solid-state electrolytes. *Energy & Environmental Science* **2024***, 17*, 6034–6045.

(35) Zhang, J.; Li, Q.; Zeng, Y.; Tang, Z.; Sun, D.; Huang, D.; Peng, Z.; Tang, Y.; Wang, H. Non-flammable ultralow concentration mixed ether electrolyte for advanced lithium metal batteries. *Energy Storage Mater.* **2022***, 51*, 660–670.

(36) Wang, Y.; Li, Z.; Xie, W.; Zhang, Q.; Hao, Z.; Zheng, C.; Hou, J.; Lu, Y.; Yan, Z.; Zhao, Q.; Chen, J. Asymmetric Solvents Regulated Crystallization-Limited Electrolytes for All-Climate Lithium Metal Batteries. *Angew. Chem. -Int. Edit.* **2024***, 63*, 202310905.

(37) Wan, H.; Xu, J.; Wang, C. Designing electrolytes and interphases for high-energy lithium batteries. *Nat Rev Chem* **2024***, 8*, 30–44.

(38) Meng, Y. S.; Srinivasan, V.; Xu, K. Designing better electrolytes. *Science* **2022***, 378*, eabq3750.

(39) Huang, B.; von Rudorff, G. F.; von Lilienfeld, O. A. The central role of density functional theory in the AI age. *Science* **2023***, 381*, 170–175.

(40) Yang, Z.; Zhao, Y.; Wang, X.; Liu, X.; Zhang, X.; Li, Y.; Lv, Q.; Chen, C. Y.; Shen, L. Scalable crystal structure relaxation using an iteration-free deep generative model with uncertainty quantification. *Nat Commun* **2024***, 15*, 8148.

(41) Ignacz, G.; Beke, A. K.; Toth, V.; Szekely, G. A Hybrid Modelling Approach to Compare Chemical Separation Technologies in Terms of Energy Consumption and Carbon Dioxide Emissions. *Nature Energy 2024 10:3* **2024**, *10* (3), 308–317.

(42) Kumar, S.; Ignacz, G.; Szekely, G. Synthesis of Covalent Organic Frameworks Using Sustainable Solvents and Machine Learning. Green Chemistry **2021**, *23* (22), 8932–8939.

(43) Mashhadimoslem, H.; Abdol, M. A.; Karimi, P.; Zanganeh, K.; Shafeen, A.; Elkamel, A.; Kamkar, M. Computational and Machine Learning Methods for $CO_2$ Capture Using Metal-Organic Frameworks. *ACS Nano* **2024**, *18* (35), 23842–23875.

(44) Yang, T.; Zhou, J.; Song, T. T.; Shen, L.; Feng, Y. P.; Yang, M. High-Throughput Identification of Exfoliable Two-Dimensional Materials with Active Basal Planes for Hydrogen Evolution. *ACS Energy Lett.* **2020***, 5*, 2313–2321.

(45) Liu, X.; Zhao, Y.; Zhang, X.; Wang, L.; Shen, J.; Zhou, M.; Shen, L. Data-Driven Discovery of Transition Metal Dichalcogenide-Based Z-Scheme Photocatalytic Heterostructures. *ACS Catal.* **2023***, 13*, 9936–9945.

(46) Aldossary, A.; Campos-Gonzalez-Angulo, J. A.; Pablo-Garcia, S.; Leong, S. X.; Rajaonson, E. M.; Thiede, L.; Tom, G.; Wang, A.; Avagliano, D.; Aspuru-Guzik, A. In Silico Chemical Experiments in the Age of AI: From Quantum Chemistry to Machine Learning and Back. *Adv. Mater.* **2024***, 36*, 2402369.

(47) Kang, Y.; Kim, J. ChatMOF: an artificial intelligence system for predicting and generating metal-organic frameworks using large language models. *Nat Commun* **2024***, 15*, 4705.

(48) Sun, D.; Wu, N.; Wen, Y.; Sun, S.; He, Y.; Huang, K.; Li, C.; Ouyang, B.; White, R.; Huang, K. Understanding Ionic Transport in Perovskite Lithium-Ion Conductor $Li_{3/8}Sr_{7/16}Ta_{3/4}Hf_{1/4}O_3$: A Neutron Diffraction and Molecular Dynamics Simulation Study. *J. Mater. Chem. A*, **2025**, Advance Article.

(49) AnonymousXGBoost(eXtreme Gradient Boosting) Documentation. (accessed Oct 31, 2024). https://xgboost.readthedocs.io/en/stable/index.html

(50) Ouyang, R.; Curtarolo, S.; Ahmetcik, E.; Scheffler, M.; Ghiringhelli, L. M. SISSO: A compressed-sensing method for identifying the best low-dimensional descriptor in an immensity of offered candidates. *Phys. Rev. Mater.* **2018***, 2*, 083802.

(51) Xie, T.; Grossman, J. C. Crystal Graph Convolutional Neural Networks for an Accurate and Interpretable Prediction of Material Properties. *Phys. Rev. Lett.* **2018***, 120*, 145301.

(52) Zhang, D.; Liu, Y.; Yang, S.; Zhu, J.; Hong, H.; Li, S.; Xiong, Q.; Huang, Z.; Wang, S.; Liu, J.; Zhi, C. Inhibiting Residual Solvent Induced Side Reactions in Vinylidene Fluoride-Based Polymer Electrolytes Enables Ultra-Stable Solid-State Lithium Metal Batteries. *Adv. Mater.* **2024***, 36*, 2401549.

(53) Dagousset, G.; François, C.; León, T.; Blanc, R.; Sansiaume-Dagousset, E.; Knochel, P. Preparation of Functionalized Lithium, Magnesium, Aluminum, Zinc, Manganese, and Indium Organometallics from Functionalized Organic Halides. *Synthesis* **2014***, 46*, 3133–3171.

(54) Pence, H. E.; Williams, A. ChemSpider: An Online Chemical Information Resource. *J. Chem. Educ.* **2010***, 87*, 1123–1124.

(55) Rysselberghe, P. V. Remarks concerning the Clausius-Mossotti Law. *J. Phys. Chem.* **1932***, 36*, 1152–1155.

(56) Burford, N.; Clyburne, J. A. C.; Wiles, J. A.; Cameron, T. S.; Robertson, K. N. Tethered Diarenes as Four-Site Donors to SbCl3. *Organometallics* **1996***, 15*, 361–364.

(57) Shen, J.; Liang, Z.; Gu, T.; Sun, Z.; Wu, Y.; Liu, X.; Liu, J.; Zhang, X.; Liu, J.; Shen, L.; Zhu, M.; Liu, J. Revisiting the unified principle for single-atom electrocatalysts in the sulfur reduction reaction: from liquid to solid-state electrolytes. *Energy & Environmental Science* **2024***, 17*, 6034–6045.

(58) Li, J.; Chen, J.; Xu, X.; Wang, Z.; Shen, J.; Sun, J.; Huang, B.; Zhao, T. Enhanced Interphase Ion Transport via Charge-Rich Space Charge Layers for Ultra-Stable Solid-State Lithium Metal Batteries. *Adv. Energy Mater.* **2024***, 15,* 2402746.

# TOC Graphic

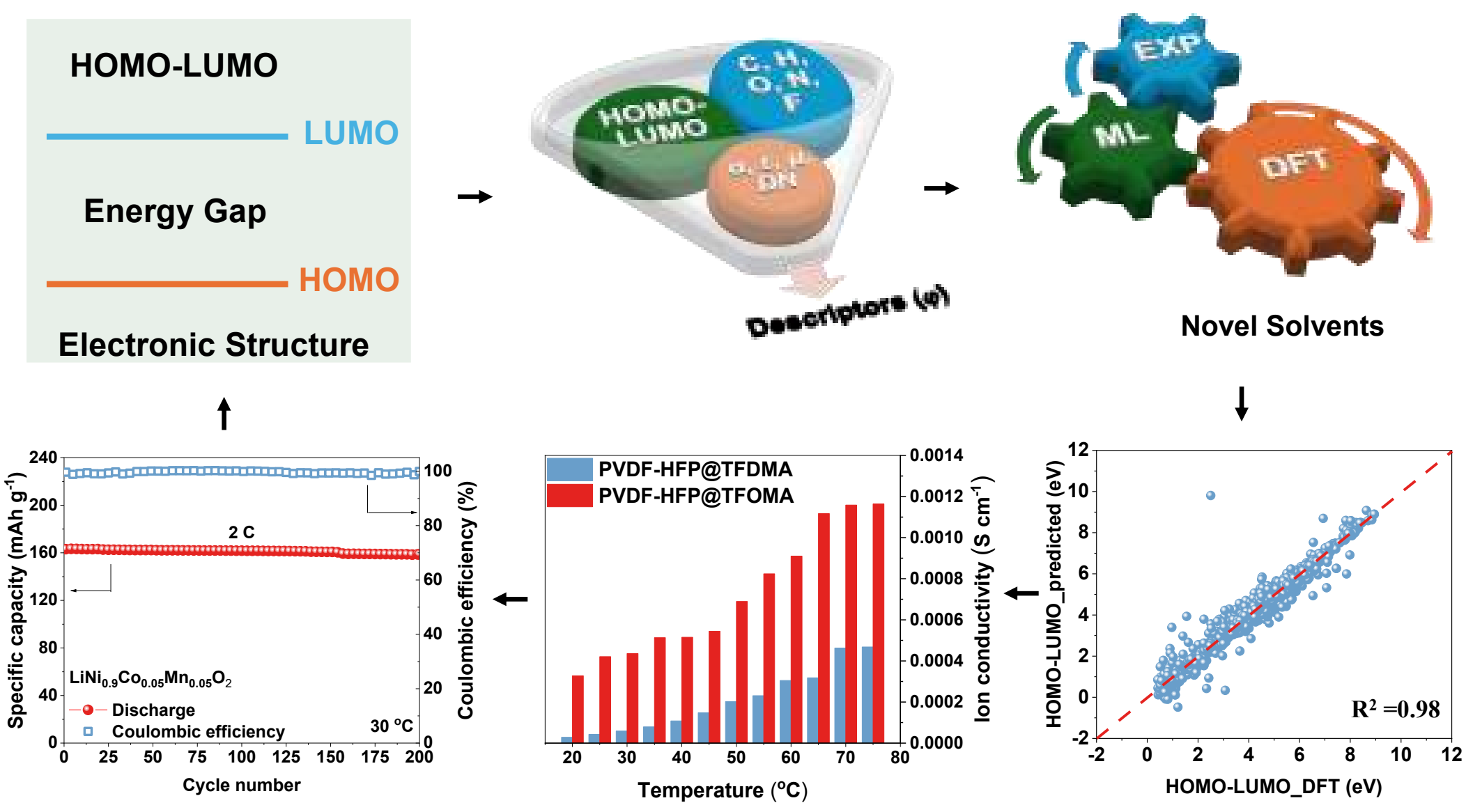

HOMO-LUMO
LUMO
Energy Gap
HOMO
Electronic Structure
HOMO-LUMO
C, H, O, N, F
Descriptors (φ)
EXP
ML
DFT
Novel Solvents
HOMO-LUMO_predicted (eV)
HOMO-LUMO_DFT (eV)
$R^2$ =0.98
PVDF-HFP@TFDMA
PVDF-HFP@TFOMA
Ion conductivity (S cm$^{-1}$)
Temperature (°C)
Specific capacity (mAh g$^{-1}$)
Coulombic efficiency (%)
2 C
$LiNi_{0.9}Co_{0.05}Mn_{0.05}O_2$
Discharge
Coulombic efficiency
30 °C
Cycle number